\documentclass[twocolumn,aps,showpacs,showkeys,
tightenlines,superscriptaddress]{revtex4}
\usepackage{epsfig,bm}

\begin{document}

\title{Reaction cross sections of carbon isotopes 
incident on a proton}

\author{B. Abu-Ibrahim}
\affiliation{RIKEN Nishina Center, RIKEN, Wako-shi, 
Saitama 351-0198, Japan}
\affiliation{Department of Physics, Cairo University, 
Giza 12613, Egypt}
\author{W. Horiuchi}
\affiliation{Graduate School of Science and Technology, 
Niigata University, Niigata 950-2181, Japan}
\author{A. Kohama}
\affiliation{RIKEN Nishina Center, RIKEN, Wako-shi, 
Saitama 351-0198, Japan}
\author{Y. Suzuki}
\affiliation{Department of Physics, 
Graduate School of Science and Technology, 
Niigata University, Niigata 950-2181, Japan}


\begin{abstract}
We systematically study total reaction cross sections 
of carbon isotopes with $N=6$-$16$ on a proton 
target for wide range of incident energies, 
putting an emphasis 
on the difference from the case of a carbon target.
The analysis includes the reaction cross sections 
of $^{19,20,22}$C at 40 $A$MeV, 
the data of which have recently been measured at RIKEN.
The Glauber theory is used to calculate the reaction cross sections. 
To describe the intrinsic structure of the carbon isotopes, 
we use a Slater determinant generated from a phenomenological 
mean-field potential, and construct the density distributions. 
To go beyond the simple mean-field model, 
we adopt two types of dynamical models: 
One is a core+$n$ model for odd-neutron nuclei, 
and the other is a core+$n$+$n$ model for $^{16}$C and $^{22}$C.
We propose empirical formulas 
which are useful in predicting unknown cross sections. 
\end{abstract}


\maketitle

\section{Introduction}
Reactions of unstable neutron-rich nuclei with a proton target 
are of current interest \cite{arkb,kohama} 
since such reactions are at present
the major means to sensitively probe the matter densities of 
exotic nuclei, especially the region of nuclear surface.
If one appropriately selects incident energies, 
protons could be more sensitive to 
neutron distributions than proton distributions of nuclei. 

The structure of carbon isotopes has recently attracted 
much attention. Several works have been done already 
experimentally \cite{imai,yamaguchi,nakamura,maddalena} 
and theoretically \cite{suzukic16,horiuchi,itagaki,kanada,sagawa}. 
For example, the structure of $^{22}$C has been
studied by two (W.H. and Y.S.) of the present authors 
in a three-body model of $^{20}$C+$n$+$n$.
They showed that it has a Borromean character \cite{hs}.

The purpose of this paper is to report 
a systematic analysis of the total reaction cross sections 
of carbon isotopes incident on a proton at 
energies from 40 $A$MeV to 800 $A$MeV, 
and predict the cross sections of neutron-rich isotopes. 
We also estimate the cross sections contributed by 
protons or neutrons in the nuclei of carbon-isotopes. 
This study is motivated by an ongoing measurement 
of the reaction cross section of $^{22}$C at RIKEN ~\cite{tanaka}. 

Recently, we have performed systematic analyses 
of total reaction cross 
sections of carbon isotopes on $^{12}$C for wide energy range 
using the Glauber model \cite{hsbk}. 
We found reasonable parameterizations of nucleon-nucleon 
scattering amplitudes, and obtained fairly good agreement 
with available data. 
We predicted the total reaction cross section of 
a neutron-rich isotope, $^{22}$C on $^{12}$C, and 
obtained a sizable effect of the extended surface.
Another purpose of this study is to discuss the advantage and 
disadvantage of a proton and a carbon target.

In this paper, we adopt the same prescription 
as our previous work \cite{hsbk} 
for describing the nuclear structure, and 
calculate the total reaction cross sections 
of proton-carbon isotopes similarly to the case of $^{12}$C target.
We calculate
systematically total reaction cross sections for wide energy range 
using the Glauber model. 
Of course, we should note that the model may not be so good 
at 40 $A$MeV.

We treat the interactions of proton-proton 
and proton-neutron separately. 
The wave functions of carbon isotopes are generated 
based on a simple mean-field model.
To go beyond that, we adopt two types of dynamical models: 
One is a core+$n$ model for an odd $N$ nucleus, and the 
other is a core+$n$+$n$ model for $^{16}$C and $^{22}$C.
The reason for the latter model is explained in Ref.~\cite{hsbk}. 
We do not take into account the Coulomb potential, 
which would affect the magnitude of the cross sections 
for the low energy processes to some extent, 
but, for the present discussion, the effect is minor.

This paper is organized as follows:
The reaction models for the calculations
of reaction cross sections 
are presented in Sec.~II. 
We explain our input data in Sec.~III.
We present the cross section calculation in Sec.~IV.
The contributions of the protons and neutrons inside 
an isotope to the reaction cross section are presented in Sec.~V.
Summary is given in Sec.~VI. 
In Appendix, we discuss the parameterization of 
the nucleon-nucleon scattering amplitude.

\section{The Glauber model for reaction cross section calculations}

Here we summarize our basic formula for the following discussions.

The total reaction cross section of proton-nucleus collisions is 
expressed as 
\begin{equation}
  \sigma_{\rm R} = \int{d{\bm b}\,
  \left(1-\big|{\rm e}^{i\chi({\bm b})}\big|^{2}\right)},
\label{reaccs}
\end{equation}
where ${\bm b}$ is the impact parameter vector
perpendicular to the beam ($z$) direction, and 
$\chi({\bm b})$ is the phase-shift function defined below.
We calculate this quantity using the Glauber theory. 

The Glauber theory provides us with an excellent framework to 
describe high energy reactions. 
In this framework, the optical phase-shift 
function (the elastic $S$-matrix) 
for proton-nucleus scattering is given by~\cite{Glauber}
\begin{eqnarray}
   {\rm e}^{i\chi({\bm b})}
   &=& \langle\psi_{0}\vert
     \prod_{i=1}^{A}
     \Big[1 - \frac{1+\tau_{3_i}}{2}
              \Gamma_{pn}({\bm b} + {\bm s}_{i})
\nonumber \\
     & & \ \ \qquad \qquad 
          -   \frac{1-\tau_{3_i}}{2}
              \Gamma_{pp}({\bm b} + {\bm s}_{i}) 
     \Big]|\psi_{0} \rangle, 
\label{opsf}
\end{eqnarray}
where $\psi_{0}$ is the intrinsic (translation-invariant) 
$A$-nucleon wave function 
of the projectile's ground state 
($A$ is the mass number of the projectile), 
and ${\bm s}_{i}$ is the projection onto the $xy$-plane 
of the nucleon coordinate relative to 
the center-of-mass of the projectile. 
Here $\tau_{3_i}$ is $1$ for neutron and $-1$ for proton.

When we apply this framework to low energy processes, 
such as the one less than 100 MeV, 
its usefulness should be carefully assessed. 
As a prescription, we carefully choose the parameters 
of the nucleon-nucleon scattering amplitude 
so as to reproduce the reaction cross sections 
of proton-$^{12}$C scatterings in consistent with 
those of $^{12}$C-$^{12}$C scatterings \cite{hsbk}. 

The profile function, $\Gamma_{pN}$, 
for $pp$ and $pn$ scatterings, is usually parameterized in the form;
\begin{equation}
  \Gamma_{pN}({\bm b}) =
  \frac{1-i\alpha_{pN}}{4\pi\beta_{pN}}\,\,
  \sigma_{pN}^{\rm tot}\, {\rm e}^{-{\bm b}^2 /(2\beta_{pN}) },
\label{gfn}
\end{equation}
where $\alpha_{pN}$ is the ratio of 
the real to the imaginary part of the 
$pp$ ($pn$) scattering amplitude in the forward direction, 
$\sigma_{pN}^{\rm tot}$ is the $pp$ ($pn$) total cross sections, 
and $\beta_{pN}$ is the slope parameter of the 
$pp$ ($pn$) elastic scattering differential cross section. 
We parameterize the nucleon-nucleon scattering amplitude 
with a single Gaussian, 
because we find that double Gaussians give numerically
almost the same reaction cross sections 
as the single Gaussian.
We discuss this point in Appendix.

There are several approximate expressions of the Glauber model
on the market. We explain some of the expressions below. 

In the optical limit approximation (OLA), 
the phase-shift function of proton-nucleus scattering
is given by
\begin{equation}
   {\rm e}^{i\chi_{\rm OLA}({\bm b})}
   = \exp \left[i\chi_{n}({\bm b})+
                i\chi_{p}({\bm b}) \right],
\label{ola}
\end{equation}
with
\begin{eqnarray}
i\chi_{p}({\bm b}) &=& -{\int{d{\bm r} }} \rho_{p}({\bm r})
                        \Gamma_{pp}({\bm s} + {\bm b}), 
\nonumber\\
i\chi_{n}({\bm b}) &=& -{\int{d{\bm r} }} \rho_{n}({\bm r})
                        \Gamma_{pn}({\bm s}+{\bm b}), 
\label{chi}
\end{eqnarray}
where $\chi_{p}$ ($\chi_{n}$) implies 
the phase shift due to the protons (neutrons) inside the nucleus. 
The function $\rho_{p}({\bm r})$ is the proton density distribution,
and $\rho_{n}({\bm r})$ is the neutron density.

In the few-body (FB) calculation, the OLA is used  
for the integration involving the coordinates of the core nucleons, 
while the integration for the valence-nucleon coordinate 
is performed without any approximation~\cite{yos,ogawa,bost,ATT96}. 
In this treatment, Eq.~(\ref{opsf}) is reduced to the following 
expression for the case of core+$n$ configuration:
\begin{equation}
  {\rm e}^{i\chi_{\rm FB}({\bm b})}
  =  \langle \varphi_{0} |
     {\rm e}^{i\chi_{{\rm C}p}({\bm b}_{\rm C}) 
            + i\chi_{pn}({\bm b}_{\rm C}+{\bm s})} 
    | \varphi_{0} \rangle, 
\label{fb}
\end{equation}
with 
\begin{equation}
{\bm b}_{\rm C} = 
{\bm b} -\textstyle{\frac{1}{{\rm A}_{\rm P}}}{\bm s}, 
\end{equation}
where $\varphi_0$ is the single-particle wave function 
of the  valence nucleon 
and ${\bm b}_{\rm C}$ is the impact parameter between 
the proton and the core. 
The phase-shift function, $\chi_{{\rm C}p}$, of 
the proton-core scattering 
is defined in exactly the same way as Eq.~(\ref{ola}). 
The proton-neutron phase-shift function, $\chi_{pn}$, 
is defined through the relation; 
$\exp(i\chi_{pn}({\bm b}))=1-\Gamma_{pn}({\bm b})$. 
In this paper, we adopt both 
Eqs.~(\ref{ola}) and (\ref{fb}) to calculate 
the phase-shift function. 

In the discussion below, 
we adopt the kinematics of the projectile's rest frame. 
We specify processes by the energy of an incident proton. 
For example, the energy of 40 $A$MeV 
of an incident nucleus of the mass number $A$ 
in the proton-fixed frame corresponds to 
the energy of 40 MeV of an incident proton 
in the projectile's rest frame.

\section{input data}
\label{input}

In this section, we list the input 
quantities in the calculations of the reaction cross sections,
and give some discussions.   

The inputs for Eq.~(\ref{opsf}) 
are the projectile's intrinsic $A$-nucleon wave function 
and the parameters of the $pp$ and $pn$ profile functions.
For Eq.~(\ref{ola}), we need only proton and 
neutron intrinsic densities and the parameters of the 
profile functions. For Eq.~(\ref{fb}), 
we need the single-particle wave function 
for the valence neutron as well.

The parameters of $\Gamma_{pp}$ and $\Gamma_{pn}$ are 
taken from Refs.~\cite{hostachy,ray}.
In Ref.~\cite{hostachy}, the experimental values of 
$\sigma_{pp}$, $\sigma_{pn}$, $\alpha_{pp}$ 
and $\alpha_{pn}$ are listed 
in energy range from 20 MeV to 300 MeV.

The parameters $\beta_{pp}$ and $\beta_{pn}$ are determined from the 
fact that the total elastic cross section, $\sigma_{pN}^{\rm el}$, 
is equal to the total 
cross section in this energy range, 
since only the elastic scattering is energetically possible 
until the pion production threshold is open.
For the profile function, Eq.~(\ref{gfn}), we have \cite{ogawa}
\begin{equation}
   \sigma_{pN}^{\rm el}=\frac{1+\alpha_{pN}^2}{16\pi \beta_{pN}}
   \left(\sigma_{pN}^{\rm tot}\right)^2.
\label{el-total}
\end{equation}
Since $\sigma_{pN}^{\rm el}=\sigma_{pN}^{\rm tot}$,
we can derive the following expression for $\beta_{pN}$;
\begin{equation}
   \beta_{pN} =
   \frac{1+\alpha_{pN}^2}{16\pi }\sigma_{pN}^{\rm tot}.
\label{beta} 
\end{equation}
The experimental data of Ref.~\cite{hostachy} 
and the $\beta_{pN}$ values determined from Eq.~(\ref{beta}) 
are listed in Table~\ref{parameter}. 
In Ref.~\cite{ray}, all the needed parameters are listed 
in the energy range from 100 MeV to 1000 MeV. 
They are also given in Table~\ref{parameter}.
For the energy higher than 300 MeV, 
$\beta_{pN}$ is determined from 
Eq.~(\ref{el-total}) using both the data of 
$\sigma_{pN}^{\rm el}$  and $\sigma_{pN}^{\rm tot}$. 
The data on $\sigma_{pN}^{\rm el}$ are taken 
from PDG tabulation~\cite{pdg} 
and the uncertainty of the data is fairly large.

\begin{table}[b]
\caption{Parameters of the $pn$ and $pp$ profile functions 
as defined in Eq.~(\ref{gfn}). $E$ is the 
projectile's incident energy.}
\label{parameter}
\begin{center}
\begin{tabular}{ccccccccc}
\hline\hline
$E$ & & $\sigma_{pp}^{\rm tot}$ & $\alpha_{pp}$ & $\beta_{pp}$ &
& $\sigma_{pn}^{\rm tot}$ & $\alpha_{pn}$ & $\beta_{pn}$ \\
(MeV) & &(fm$^2$) &  &  (fm$^2$) && (fm$^2$) &  &  (fm$^2$)\\
\hline
 40 & &7.0  &  1.328 & 0.385 & &21.8 & 0.493 & 0.539 \\
 60 & &4.7  &  1.626 & 0.341 & &13.6 & 0.719 & 0.410 \\
 80 & &3.69 & 1.783  & 0.307 & &9.89 & 0.864 & 0.344 \\
100 & &3.16 & 1.808  & 0.268 & &7.87 & 0.933 & 0.293 \\
120 & &2.85 & 1.754  & 0.231 & &6.63 & 0.94  & 0.248 \\
140 & &2.65 & 1.644  & 0.195 & &5.82 & 0.902 & 0.210 \\
160 & &2.52 & 1.509  & 0.164 & &5.26 & 0.856 & 0.181 \\
180 & &2.43 & 1.365  & 0.138 & &4.85 & 0.77  & 0.154 \\
200 & &2.36 & 1.221  & 0.117 & &4.54 & 0.701 & 0.135 \\
240 & &2.28 & 0.944  & 0.086 & &4.13 & 0.541 & 0.106 \\
300 & &2.42 & 0.626  & 0.067 & &3.7  & 0.326 & 0.081 \\
425 & &2.7  & 0.47   & 0.078 & &3.32 & 0.25  & 0.0702 \\
550 & &3.44 & 0.32   & 0.11  & &3.5  &$-0.24$  & 0.0859 \\
650 & &4.13 & 0.16   & 0.148 & &3.74 &$-0.35$ & 0.112 \\
700 & &4.43 & 0.1    & 0.16  & &3.77 &$-0.38$  & 0.12  \\ 
800 & &4.59 & 0.06   & 0.185 & &3.88 &$-0.2$   & 0.12  \\
1000& &4.63 & $-$0.09  & 0.193 && 3.88 &$-$0.46& 0.151  \\
\hline\hline
\end{tabular}
\end{center}
\end{table}

As we show in Appendix, 
the description of the $pn$ elastic 
differential cross section with these parameters is reasonable, 
but not perfect especially in the forward direction. 
Fortunately, this does not affect 
the total reaction cross sections. 
We discuss it in some detail in the appendix. 

\begin{figure*}[p]
\epsfig{file=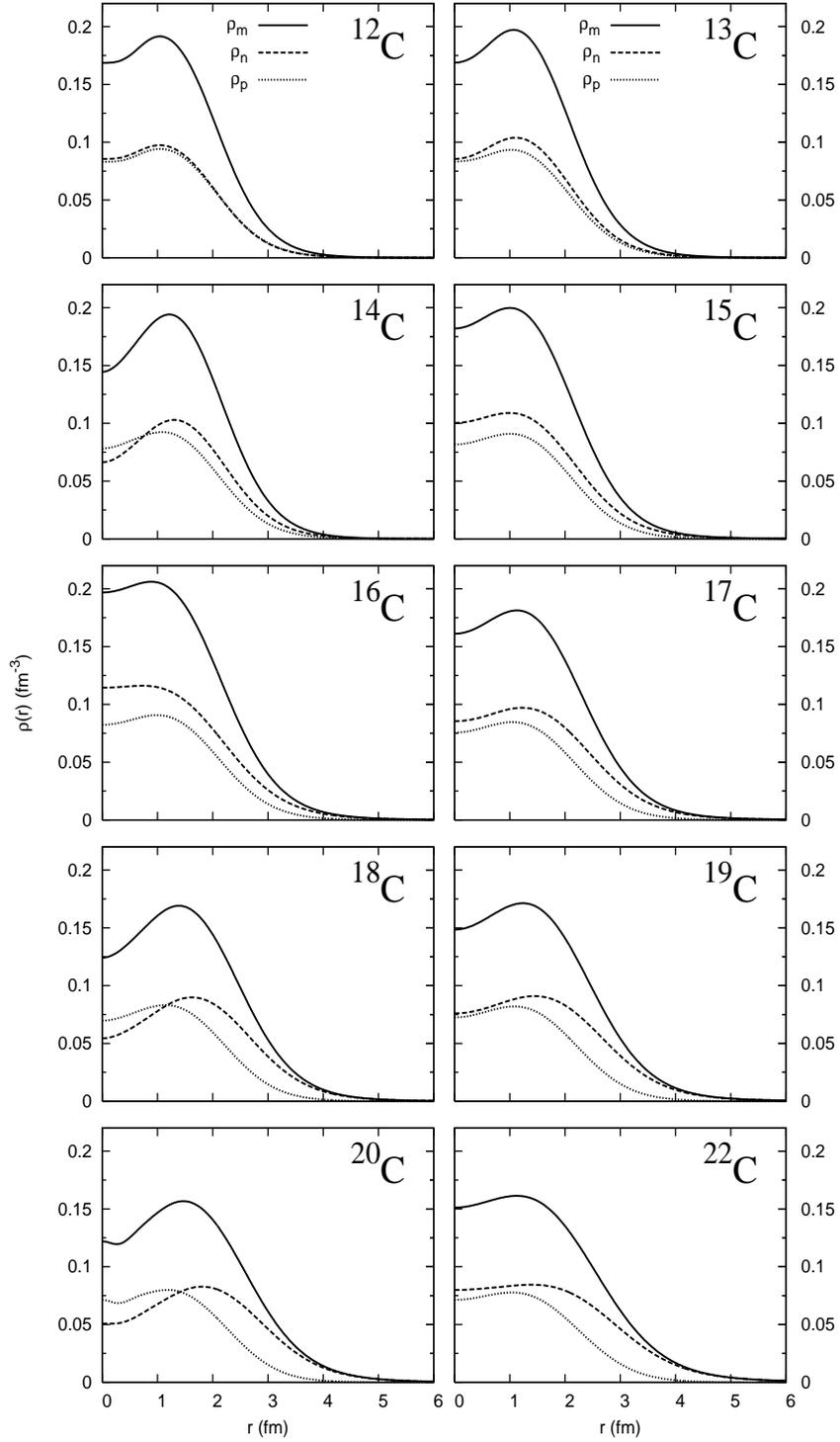,width=12.0cm,height=20.22cm}
\caption{Density distributions of the carbon isotopes.
The dotted curve shows the proton density, the dashed curve 
the neutron density, and the solid curve the matter density. }
\label{cdens}
\end{figure*}

The densities that we use here are constructed from 
a core+$n$ model 
for the odd isotopes, $^{13,15,17,19}$C, 
where the cores are $^{12,14,16,18}$C, respectively.
For $^{16,22}$C, a core+2$n$ model is assumed.
The densities of the carbon isotopes are displayed in 
Fig.~\ref{cdens}, and the corresponding root-mean-square 
(rms) radii 
are summarized in Table~\ref{radii}. 
The detail of these densities can be found in Ref.~\cite{hsbk}.

\begin{table}[b]
\caption{The rms radii in fm
of matter, neutron and proton density distributions
for the carbon isotopes.}
\label{radii}
\begin{center}
\begin{tabular}{c|ccc}
\hline\hline
Isotopes  & $r_m$ & $r_n$ & $r_p$  \\
\hline
$^{12}$C &  2.31 &  2.30 &  2.33 \\
$^{13}$C &  2.37 &  2.40 &  2.34  \\       
$^{14}$C &  2.39 &  2.46 &  2.31 \\       
$^{15}$C &  2.65 &  2.84 &  2.34 \\       
$^{16}$C &  2.66 &  2.83 &  2.34 \\      
$^{17}$C &  2.94 &  3.20 &  2.38  \\      
$^{18}$C &  2.78 &  2.96 &  2.36  \\       
$^{19}$C &  3.09 &  3.37 &  2.38 \\       
$^{20}$C &  2.99 &  3.23 &  2.37 \\       
$^{22}$C &  3.58 &  3.92 &  2.43 \\       
\hline\hline
\end{tabular}
\end{center}
\end{table}

\section{Prediction of the reaction cross sections}

Here we show our numerical results of the total reaction 
cross sections of proton-carbon isotopes reactions. 

\begin{figure}[t]
\epsfig{file=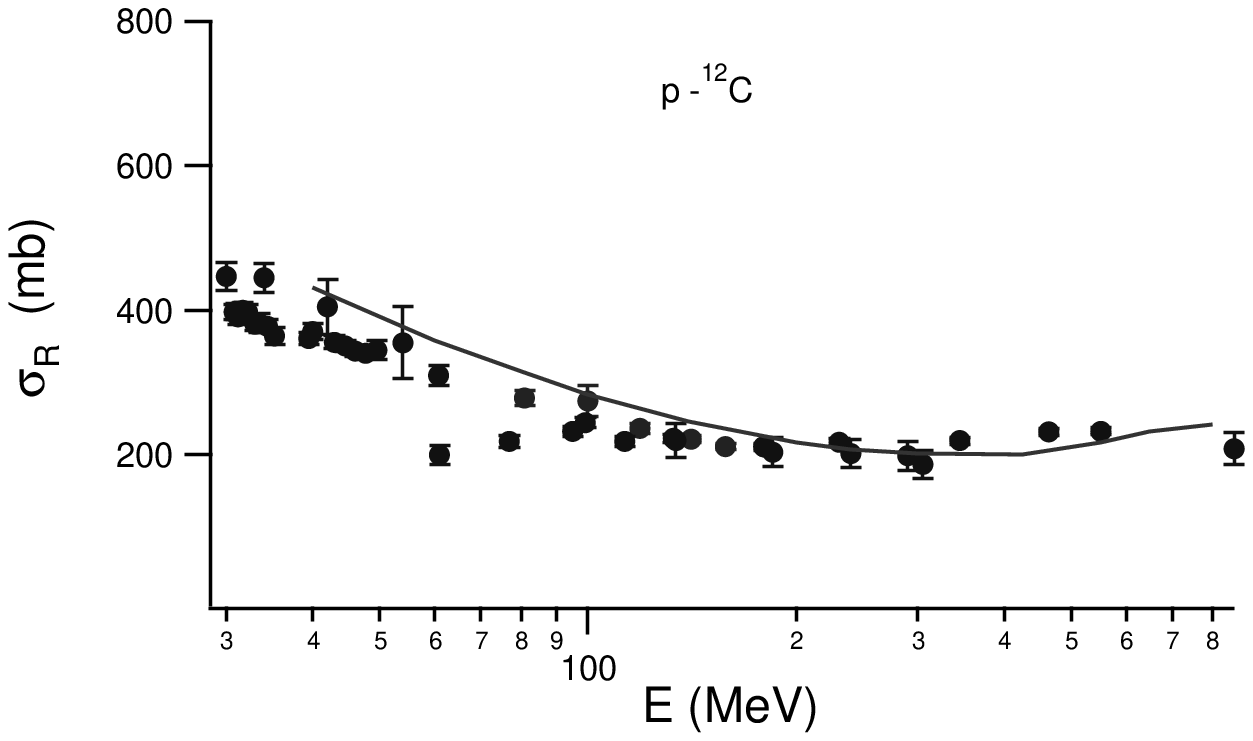,width=9.0cm,height=6.cm}
\caption{Comparison of the numerical results 
with the experimental data for the total
reaction cross sections of proton-$^{12}$C reaction
as a function of energy. 
The experimental data are taken from Refs.~\cite{carlson,auce}.
}
\label{p12c-1}
\end{figure}

Before we predict the reaction cross sections for the isotopes, 
we first show how well our densities and the parameters 
of the profile functions 
fit the experimental data of the proton-$^{12}$C total reaction 
cross sections.
In Fig.~\ref{p12c-1}, we compare the numerical results 
with the experimental data 
over the energy range from 40 MeV to 800 MeV.
As one can see from the figure, 
they reasonably agree with each other over all the energy range. 
At energies lower than 100 MeV, 
where the data fluctuate by 15 $\%$ at most, 
our results follow the largest data. 
In this energy region, a systematic uncertainty of 
our approach is estimated to be about 15 $\%$, 
which is consistent with the estimation 
by two of us (B.A.-I. and Y.S.) 
for the case of $^6$He+$^{12}$C reaction 40 $A$MeV \cite{baiys}.
They confirmed that the eikonal approximation gives 
about 15 $\%$ larger cross sections 
than those by the quantum-mechanical (exact) calculation.

Now we show our predictions for all the carbon 
isotopes at all the energies 
using the parameters given in Table~\ref{parameter}.
The numerical results of the total reaction cross sections
are summarized in Table~\ref{rxs}.

\begin{table}[t]
\caption{Total reaction cross sections of proton-carbon isotopes 
in units of mb. $E$ is the projectile's incident energy.
}
\label{rxs}
\begin{center}
\begin{tabular}{c|cccccccccc}
\hline\hline
$E$      &\multicolumn{10}{c}{Isotopes}\\
(MeV)& $^{12}$C & $^{13}$C & $^{14}$C &
          $^{15}$C & $^{16}$C & $^{17}$C &
          $^{18}$C & $^{19}$C & $^{20}$C & $^{22}$C\\
 
\hline
  40 & 432 & 467 & 489 & 580 & 605 & 682 & 662 & 758 & 761 & 957 \\
 100 & 284 & 308 & 327 & 372 & 394 & 436 & 443 & 491 & 509 & 604 \\
 200 & 218 & 236 & 252 & 282 & 300 & 330 & 340 & 372 & 390 & 453 \\
 300 & 202 & 218 & 231 & 257 & 273 & 299 & 309 & 337 & 353 & 407 \\
 425 & 200 & 214 & 227 & 251 & 265 & 289 & 300 & 324 & 339 & 389 \\
 550 & 217 & 231 & 242 & 267 & 282 & 307 & 315 & 341 & 356 & 408 \\
 650 & 233 & 247 & 259 & 284 & 299 & 324 & 332 & 359 & 374 & 429 \\
 800 & 243 & 257 & 268 & 294 & 309 & 335 & 342 & 373 & 385 & 442 \\
\hline\hline
\end{tabular}
\end{center}
\end{table}

Let us estimate the contributions of the breakup effect
although it is expected to be small for a proton target.
Equations~(\ref{opsf}) and (\ref{fb}) contain the breakup 
effect, while Eq.~(\ref{ola}) does not.
We compare them to estimate the breakup effect.
As an illustrative example, we calculate the 
reaction cross section of a typical halo nucleus, $^{19}$C, 
incident on a proton using Eqs.~(\ref{ola}) and (\ref{fb}).
We assume the structure of $^{19}$C as $^{18}$C+$n$ 
with the one-neutron separation energy of 0.581 MeV~\cite{NP}.  
The numerical results  
at 40 MeV and 800 MeV are 763 mb (758 mb) and 
372 mb (373 mb) respectively when 
Eq.~(\ref{fb}) (Eq.~(\ref{ola})) is used. 
The difference is less than one $\%$, 
which is consistent with the results of Ref.~\cite{yos}.     
The breakup effect can therefore be neglected. 
This validates our use of Eq.~(\ref{ola}). 

For convenience, we introduce the black-sphere radius, $a$, 
defined through \cite{BS2} 
\begin{equation}
  \sigma_{\rm R} = \pi a^{2}.
\label{bsr}
\end{equation}
Following the Carlson's prescription~\cite{carlson},
we fit the numerical results 
by parameterizing the radius, 
$a$, using a simple geometric picture with a correction term;  
\begin{equation}
   a = C_0 + r_{0}A^{1/3}. 
\label{carl}
\end{equation}
This includes a $A^{1/3}$ correction 
in addition to the simple geometrical $A^{2/3}$ term. 
In Ref.~\cite{carlson}, 
Carlson used $R_{p}$ instead of $C_0$ here.
He fitted the reaction cross sections of stable nuclei incident on 
a proton target in the energy range from 40 MeV to 560 MeV.

\begin{figure}[t]
\epsfig{file=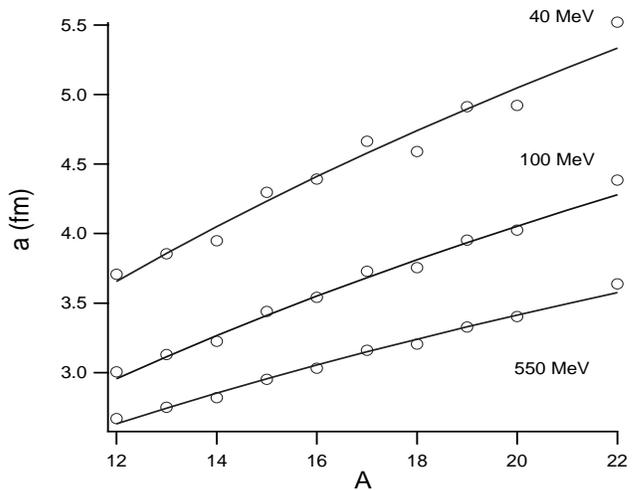,width=9cm,height=6.75cm}
\caption{Comparison of the black square radii, 
$a=\sqrt{\sigma_{\rm R}/\pi}$, 
for the numerical results of the carbon isotopes (open circles) 
with the fit using Eq.~(\ref{carl}) (solid line)
as a function of the mass number. 
}
\label{car-reac}
\end{figure}

In Fig.~\ref{car-reac}, 
we compare our numerical results (open circles) 
listed in Table~\ref{rxs}
with the fit using Eq.~(\ref{carl}) (solid lines) 
at 40, 100 and 550 MeV. 
The values of $C_0$ and $r_0$ extracted from the fit
are given in Table~\ref{tbl-car}.    
These values that we find are different 
from those obtained by Carlson, 
which are given in the parentheses in the table. 

The parameter $C_0$ implies the strength 
of $A^{1/3}$ correction 
to $A^{2/3}$-dependence of the reaction cross sections. 
The values in Table~\ref{tbl-car} decrease with the energy, 
which is consistent with the geometrical picture 
of the cross section at high energy, 
because $\sigma_{\rm R} \propto A^{2/3}$ for proton-nucleus reaction.

As one can see from Fig.~\ref{car-reac}, 
the curves with our parameters 
nicely fits the numerical results for the stable isotopes
as well as the neutron-rich unstable isotopes.
The estimations with Carlson's parameters underestimates 
our numerical results for neutron-rich cases. 
This reflects an anomalous mass number dependence of the size 
of such exotic nuclei.
Even with the new parameterization, the reaction 
cross section of $^{22}$C is even larger than the fit, 
especially at 40 MeV. 
This would suggest an extended surface structure of $^{22}$C. 
We believe that this simple fitting formula will serve 
as a reference 
for discussions of the total reaction cross sections. 

Moreover, we empirically deduce the rms nuclear matter radii, 
using the black-sphere radius, $a$, of Eq.~(\ref{bsr}). 
If we assume a rectangular density distribution for nuclei, 
we obtain 
\begin{equation}
  r_{\rm BS} = \sqrt{\frac{3}{5}}a
  = \sqrt{\frac{3}{5}} \sqrt{\frac{\sigma_{\rm R}}{\pi}}.
\label{rbs}
\end{equation}
This radius clearly depends on the incident energy.
At around 100 MeV, this $r_{\rm BS}$ value happens to agree 
reasonably well with $r_m$ listed in Table~\ref{radii}. 
This suggests that we may empirically access to the rms 
nuclear matter radii of carbon isotopes 
just by measuring $\sigma_{\rm R}$ at 100 MeV. 
This is consistent with the estimations in Ref.~\cite{BS2}. 
The authors of Ref.~\cite{BS2} pointed out that, 
for $T_p\gtrsim800$ MeV, $r_{\rm BS}$ almost completely 
agrees with the empirically deduced values 
of the rms matter radius 
for stable nuclei having mass $A\gtrsim50$,   
while it systematically deviates from the deduced values 
for $A\lesssim50$~\cite{BS2}. 
Since carbon isotopes belong to light nuclei, 
we may choose the energy which gives a little bit larger 
$\sigma_{\rm R}$ to obtain $r_{\rm BS}$ close to $r_m$


\begin{table}[b]
\caption{The parameters of Eq.~(\ref{carl}) 
which give the lines plotted in Fig.~\ref{car-reac}. 
The values in the parentheses are those of Carlson~\cite{carlson}.}
\label{tbl-car}
\begin{center}
\begin{tabular}{ccc}
\hline\hline
$E$ (MeV) & $C_0$ (fm) & $r_0$ (fm) \\
\hline
40  & $-$3.83 (1.00)    & 3.27  (1.21) \\
80  & $-$3.123          & 2.73         \\
100 & $-$2.95 ($-$0.31) & 2.58  (1.37) \\
140 & $-$2.68           & 2.38         \\
200 & $-$2.46           & 2.21         \\
240 & $-$2.36           & 2.14         \\
300 & $-$2.14           & 2.03         \\
425 & $-$1.62           & 1.85         \\
550 & $-$1.58 ($-$0.30) & 1.84  (1.33) \\
800 & $-$1.31           & 1.782        \\
\hline\hline
\end{tabular}
\end{center}
\end{table}

\begin{figure*}[t]
\epsfig{file=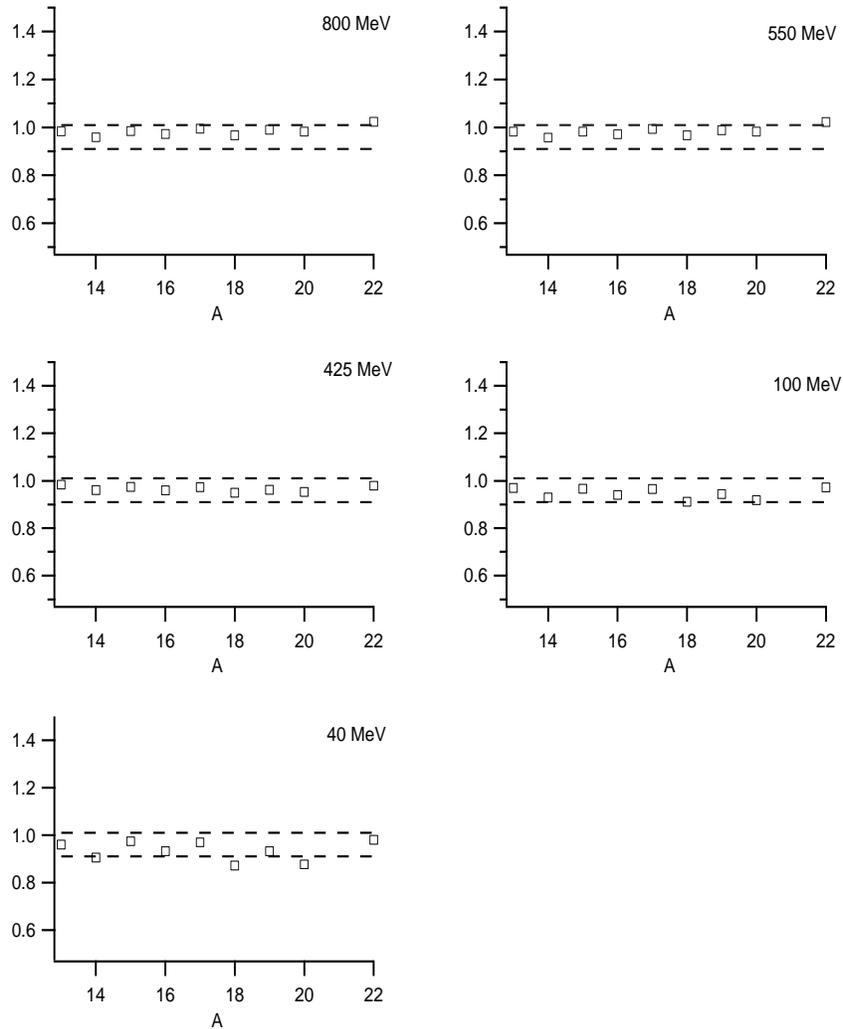,width=12cm,height=14.12cm}
\caption{The values of $R({\rm C})$ of Eq.~(\ref{emp}) 
as a function of the mass number, $A$.
The dashed lines denote 0.96$\pm$0.05 values.}
\label{emp1}
\end{figure*}

We propose another empirical formula. 
For all the carbon isotopes, we find that the 
following relation is satisfied over all the energy range: 
\begin{equation}  
   \frac{\sigma_{\rm R}(p + {_{\quad 6}^{6\!+\!N}{\rm C}})}
        {\sigma_{\rm R}(p + {^{12}_{\ \, 6}{\rm C}})}
   = R({\rm C}) \frac{6 \sigma_{pp}^{\rm tot} 
                    + N \sigma_{pn}^{\rm tot}}
                     {6 \sigma_{pp}^{\rm tot} 
                    + 6 \sigma_{pn}^{\rm tot}},
\label{emp} 
\end{equation}
with $R({\rm C}) = 0.96 \pm 0.05$.
Here $N \ge 7$ and 
$\sigma_{pp}^{\rm tot}$($\sigma_{pn}^{\rm tot}$) is 
the proton-proton 
(proton-neutron) total cross section at a given energy. 
The value of $R({\rm C})$ is obtained by averaging 
the 153 numerical results 
(9 isotopes times 17 energy points) 
of the reaction cross sections, and 
0.05 is the standard deviation of these points. 
At high energy, $R({\rm C})$
of Eq.~({\ref{emp}}) is very close to unity.
In Fig.~\ref{emp1}, we plot $R({\rm C})$
for selected energies.  
Only at 40 MeV, some points come slightly below this relation,
which would suggest the breakdown of the approximations, 
such as the fixed-scatterer approximation, 
contained in the Glauber model. 

At least for carbon isotopes, the expression~(\ref{emp})
indicates that 
if we know the reaction cross section of a stable isotope, 
we can predict the reaction cross section of other isotope 
within the error bar. 
Whether this holds for any nuclides or not
is left for a future study.

\begin{figure}[t]
\epsfig{file=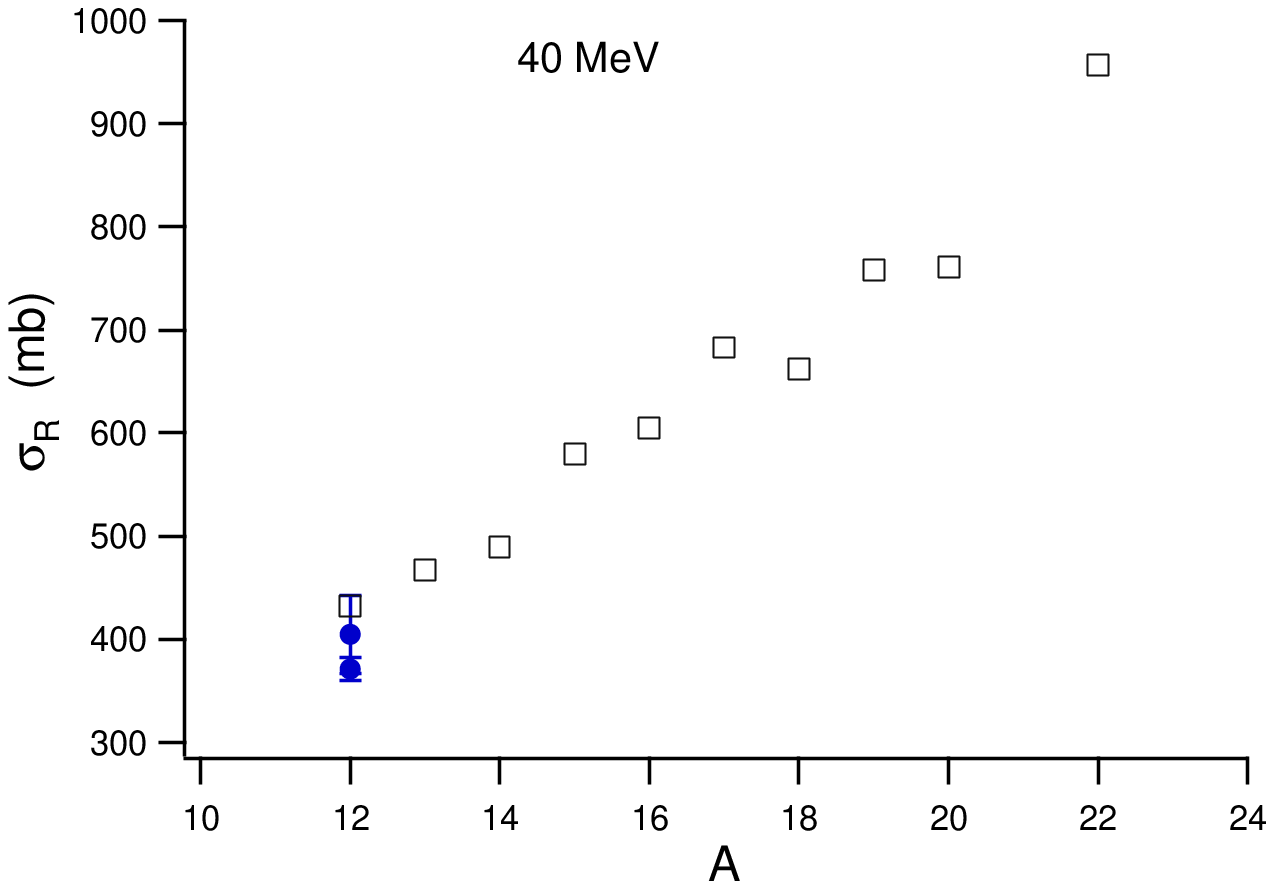,width=8.5cm,height=5.95cm}
\caption{Reaction cross section for the carbon isotopes at 40 MeV. 
The experimental data are taken from Ref.~\cite{carlson}. 
The larger one is natural carbon at 42 MeV, 
and the smaller one is $^{12}$C at 40 MeV.
The energy is converted to the case of a proton target.
The preliminary data for $^{22}$C is about 1000 mb 
with a large uncertainty~\cite{tanaka}.
}
\label{r-ciso}
\end{figure}

Experimental data are expected to appear 
at around 40 MeV for $^{19,20,22}$C~\cite{tanaka}.  
Here we predict them. 
In Fig.~\ref{r-ciso}, we compare our prediction 
for the reaction cross sections of carbon isotopes 
with the available experimental data.  
The preliminary data of proton-$^{22}$C reaction cross section 
has been reported 
to be around 1000 mb with a large uncertainty~\cite{tanaka}, 
which is consistent with our prediction. 

For $^{22}$C, we generate several densities that give 
different two-neutron separation energies of 
0.489, 0.361, 0.232 and 0.122 MeV for the last two neutrons. 
All of them lie within the error bar of the 
experimental value, $0.423\pm$1.140 MeV \cite{NP}. 
Using these densities, we calculate the reaction cross sections 
for proton-$^{22}$C reaction at 40 MeV in order to examine 
the separation-energy dependence. 
The results are listed in Table~\ref{rcs22c}. 
The change in radius from 3.6 to 4.1 fm gives 
change in the reaction 
cross section of about 50 mb at 40 MeV. 
At 800 MeV, the change in the reaction cross section 
is about 10 mb.
This gives an estimate of an uncertainty of our calculations.

\begin{table}[b]
\caption{The reaction cross sections of $^{22}$C incident 
on a proton target at 40 $A$MeV for different two-neutron 
separation energies, $S_{2n}$. 
The $r_m$ value denotes the rms matter radius.}
\label{rcs22c}
\begin{center}
\begin{tabular}{ccc}
\hline\hline
$S_{2n}$ (MeV) & $r_m$ (fm) & $\sigma_{\rm R}$ (mb)\\
\hline
0.489 & 3.6 &  957  \\
0.361 & 3.7 &  969  \\       
0.232 & 3.8 &  985  \\
0.122 & 4.1 & 1005  \\           
\hline\hline
\end{tabular}
\end{center}
\end{table}

\section{The neutron contribution to the reaction cross section}

Here we estimate 
that contribution to reaction cross sections which comes from 
the neutrons in the nucleus.  
Reactions with a proton are superior to those with a $^{12}$C
when we look into such separate contributions, because  
$^{12}$C is equally sensitive to protons and neutrons. 
For the purpose of discussion here, 
we use the following relation: 
\begin{eqnarray}
 1 - \big|{\rm e}^{i\chi_n({\bm b})+ 
                   i\chi_p({\bm b})}\big|^{2} 
&=&  \big|{\rm e}^{i\chi_p({\bm b})}\big|^{2}
     \left(1 - \big|{\rm e}^{i\chi_n({\bm b})}\big|^{2}\right)
\nonumber\\
&+&  \big|{\rm e}^{i\chi_n({\bm b})}\big|^{2}
     \left(1 - \big|{\rm e}^{i\chi_p({\bm b})}\big|^{2}\right)
\\
&+&  \left(1 - \big|{\rm e}^{i\chi_n({\bm b})}\big|^{2}\right)
     \left(1 - \big|{\rm e}^{i\chi_p({\bm b})}\big|^{2}\right).
\nonumber
\end{eqnarray}
Then we define the proton-nucleus reaction probability $P_A(b)$ and 
its decomposition into neutron and proton contributions, 
$P_n(b)$ and $P_p(b)$, as
\begin{equation}
  P_A(b) = P_n(b) + P_p(b), 
\label{defofpa}
\end{equation}
where
\begin{eqnarray}
  P_{n}(b) 
  &=& \Big|{\rm e}^{i\chi_{p}({\bm b})}\Big|^{2}
      \left(1 - \Big|{\rm e}^{i\chi_{n}({\bm b})}\Big|^{2}\right)
\nonumber\\
  &+& c \left(1 - \Big|{\rm e}^{i\chi_{n}({\bm b})}\Big|^{2}
        \right)
        \left(1 - \Big|{\rm e}^{i\chi_{p}({\bm b})}\Big|^{2}
        \right), 
\nonumber \\
   P_{p}(b)
   &=& \big|{\rm e}^{i\chi_{n}({\bm b})}\big|^{2} 
      \left(1 - \big|{\rm e}^{i\chi_{p}({\bm b})}\big|^{2}\right)
\nonumber\\
   &+& d \left(1 - \big|{\rm e}^{i\chi_{n}({\bm b})}\big|^{2}
         \right)
         \left(1 - \big|{\rm e}^{i\chi_{p}({\bm b})}\big|^{2}
         \right), 
\label{rprobf}
\end{eqnarray}
where $c + d = 1$, and $c$ and $d$ represent the neutron 
and proton contributions from the interference term, respectively. 
Equation~(\ref{reaccs}) is expressed as
\begin{equation}
   \sigma_{\rm R} = 2 \pi \int_0^{\infty} b db \, P_A(b). 
   \label{probrel}
\end{equation}

The values of $c$ and $d$ must satisfy the condition, 
$c + d = 1$, but the choice of them is not unique. 
Here we discuss two cases; 
1) $c = d = 1/2$, 2) $c = N/A$ and $d = Z/A$, 
to see the dependence of the choice. 
The former implies that 
both neutrons and protons contribute equally 
to the interference term, while 
the latter implies that 
the contribution of the neutrons and protons 
to the interference term is proportional to their numbers.

\begin{figure}[t]
\epsfig{file=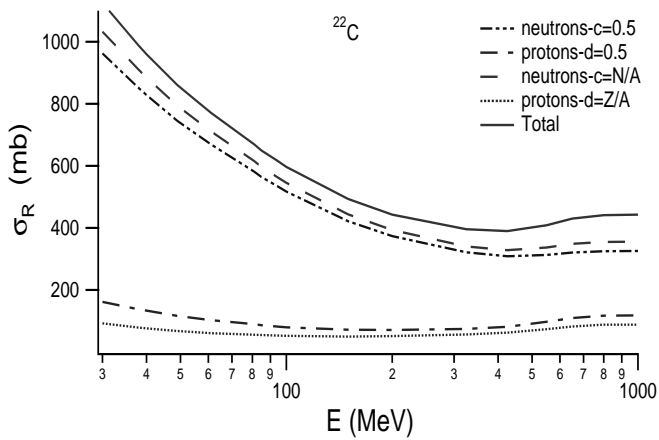,width=9cm,height=6.cm}
\caption{The total reaction cross section of proton-$^{22}$C as 
a function of energy (solid curve), and
its decomposition to the neutron and proton contributions.
} 
\label{22c-reac}
\end{figure}

In Fig.~\ref{22c-reac}, we show our predictions of 
the total reaction cross section of proton-$^{22}$C reaction 
as a function of energy (solid curve). 
We also draw the neutron and proton contributions
for $c = d = 1/2$ (dash-dot-dotted and dash-dotted) and 
$c = N/A$, $d = Z/A$ (dashed and dotted), respectively. 
Due to the fact that $^{22}$C is very neutron-rich, 
we learn from this figure that the neutron contribution 
dominates the reaction cross sections. 
Also we find that the proton and neutron contributions 
depend modestly on the choice of the values of $c$ and $d$.
For example, for $c = d = 1/2$, 
the neutron contribution to the total 
reaction cross section of $^{22}$C, 
$2 \pi \int_0^{\infty} b db \, P_n(b)
/ 2 \pi \int_0^{\infty} b db \, P_A(b)$, 
is about 0.87 and 0.73  
at 40 MeV and 800 MeV, respectively.
For $c = N/A$, $d = Z/A$, the neutron contribution 
to the total reaction cross sections is 
about 0.93 and 0.80 at 40 MeV and 800 MeV, respectively.

According to Eq.~(\ref{emp}), 
the neutron contribution to the total reaction 
cross section would be similar to 
the ratio $N\sigma_{pn}^{\rm tot}
/(Z\sigma_{pp}^{\rm tot} + N\sigma_{pn}^{\rm tot})$.     
At 40 MeV and 800 MeV, the ratio of 
$N\sigma_{pn}^{\rm tot}
/(Z\sigma_{pp}^{\rm tot} + N\sigma_{pn}^{\rm tot})$ 
reads 0.89 and 0.69, respectively.   
These values are quite similar to the above ratios 
of our numerical results.

\begin{figure*}[t]
\epsfig{file=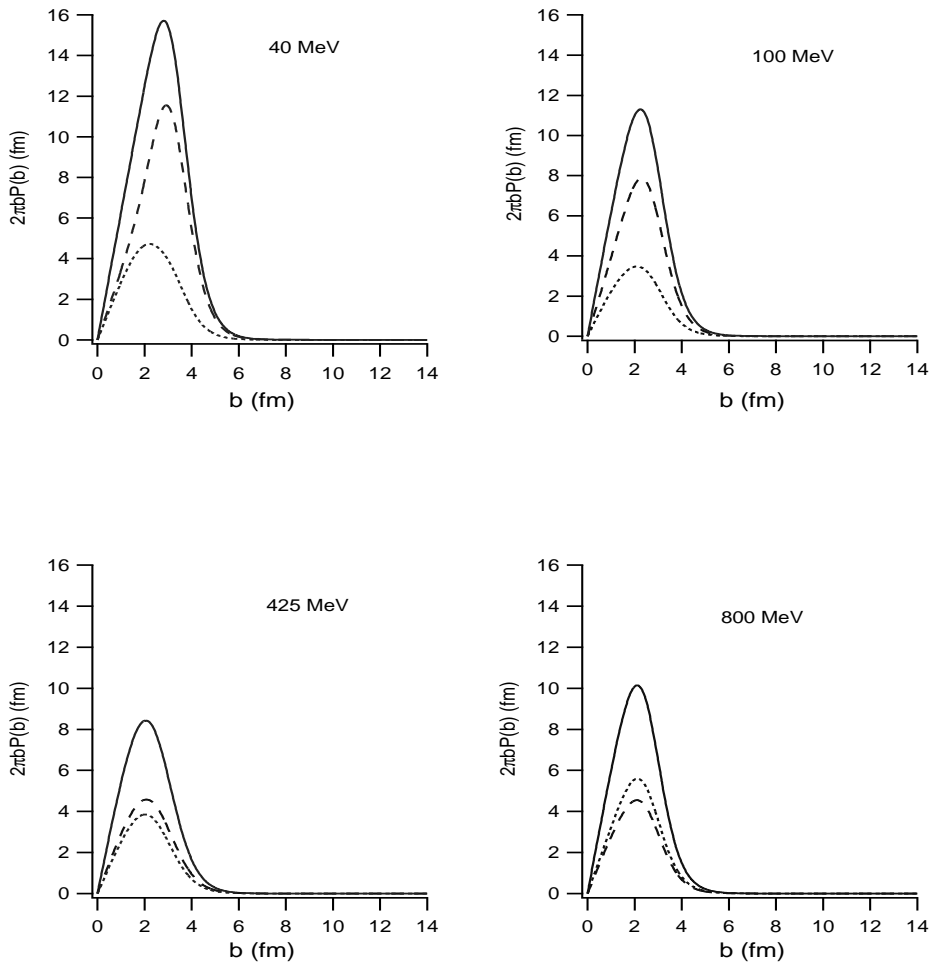,width=13cm,height=13.cm}
\caption{Reaction probability, Eq.~(\ref{rprobf}), times $2\pi b$
for proton-$^{12}$C. 
The solid curve is the total probability. 
The dashed curve is the neutron contribution, 
while the dotted curve is the proton contribution.
The choice of $c = d = 1/2$ is made.
}
\label{12cprob-f}
\end{figure*}

Figure~\ref{12cprob-f} shows the reaction 
probability  times $2\pi b$
of proton-$^{12}$C reaction 
as a function of the impact parameter, $b$. 
We plot $2\pi b P(b)$, because this quantity more directly 
reflects the contribution to $\sigma_{\rm R}$ than $P(b)$ itself. 
The solid curve represents the total 
reaction probability, $P_A(b)$, in Eq.~(\ref{defofpa}). 
The neutron contribution $P_{n}(b)$ 
is shown by the dashed curve and 
the proton contribution $P_{p}(b)$ 
is shown by the dotted curve. 
Here we draw only the case of $c = d = 1/2$. 
One can see from the figure that, at 40 MeV, 
the neutron contribution 
to the total reaction probability 
is about two times of the proton contribution, 
while at 800 MeV 
the proton contribution exceeds that of the neutron.
This reflects the behavior that the $pn$ total cross 
section, $\sigma^{\rm tot}_{pn}$, is significantly 
larger than the $pp$ 
total cross section, $\sigma^{\rm tot}_{pp}$, 
at low energy region.

\begin{figure*}
\epsfig{file=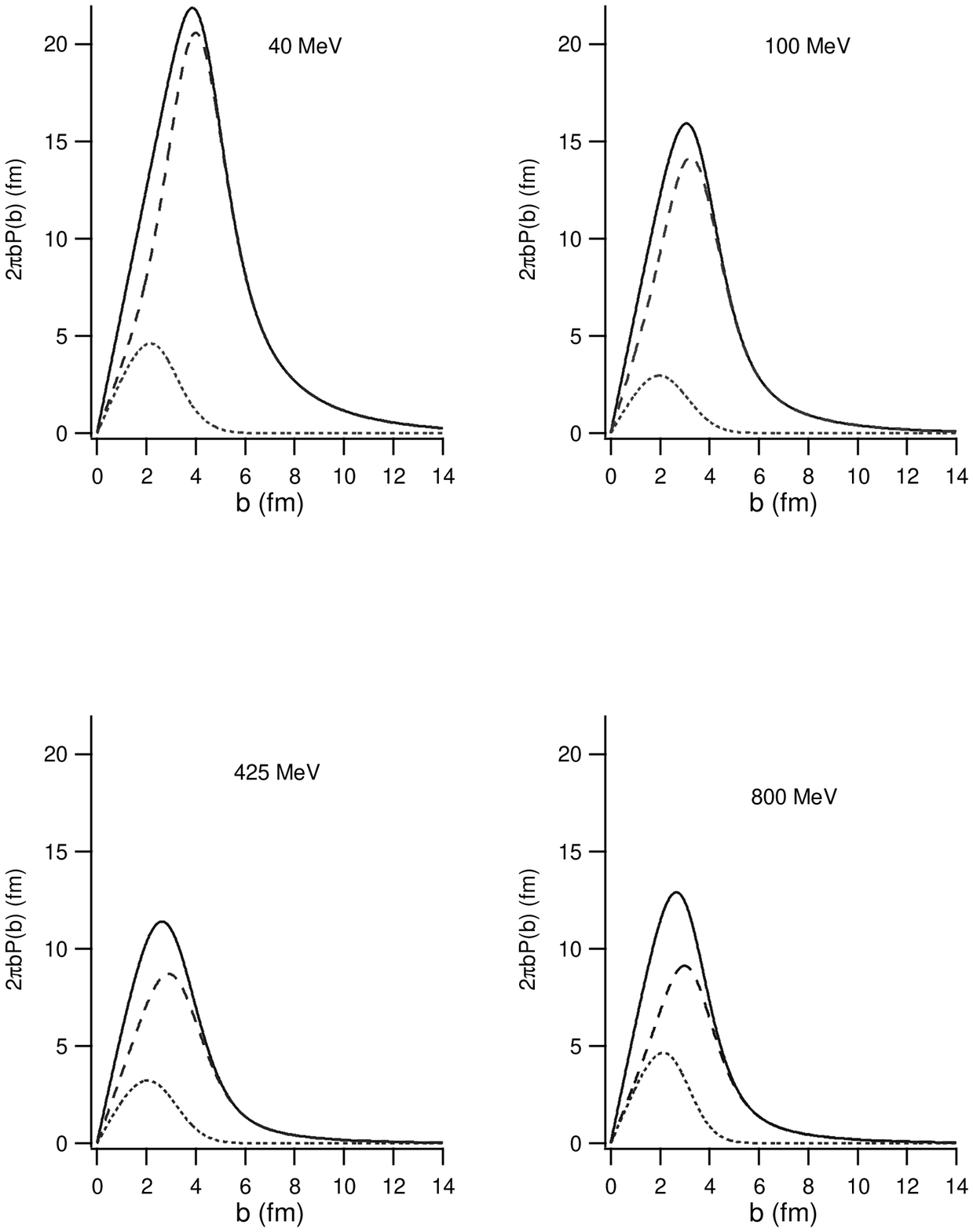,width=13cm,height=13cm}
\caption{The same as in Fig.~\ref{12cprob-f} 
but for proton-$^{22}$C case. The choice of $c = d = 1/2$ is made.
}
\label{22cprob}
\end{figure*}

Figure~\ref{22cprob} displays the reaction 
probability of proton-$^{22}$C reaction, similarly to that of 
proton-$^{12}$C reaction. In contrast to the case of $^{12}$C,  
the neutron contribution to the reaction probability is 
larger than that of the proton over all the energy range.
Also, the reaction probability on the surface region 
comes mostly from the neutron contribution 
at all the energy range. 
This is due to the large extension of the neutron density, 
as shown in Fig. 1.

The difference between the reaction probability of 
proton-$^{12}$C and that of proton-$^{22}$C is as follows: 
Let the probing position of the proton be an impact parameter 
at which $2 \pi b P(b)$ becomes a maximum. 
The probing position for the case of 
proton-$^{12}$C at 800 MeV is at 2.0 fm, and the maximum 
height is about 10 fm, 
while they are about 2.7 fm and 13 fm in the case of $^{22}$C. 
The reaction probability of proton-$^{12}$C 
at 800 MeV reaches zero at about 6 fm, 
while in the case of proton-$^{22}$C it reaches zero at about 10 fm. 
The major contribution comes from the region around the 
probing point, {\it i.e.}, the surface. 

\begin{figure*}[t]
\epsfig{file=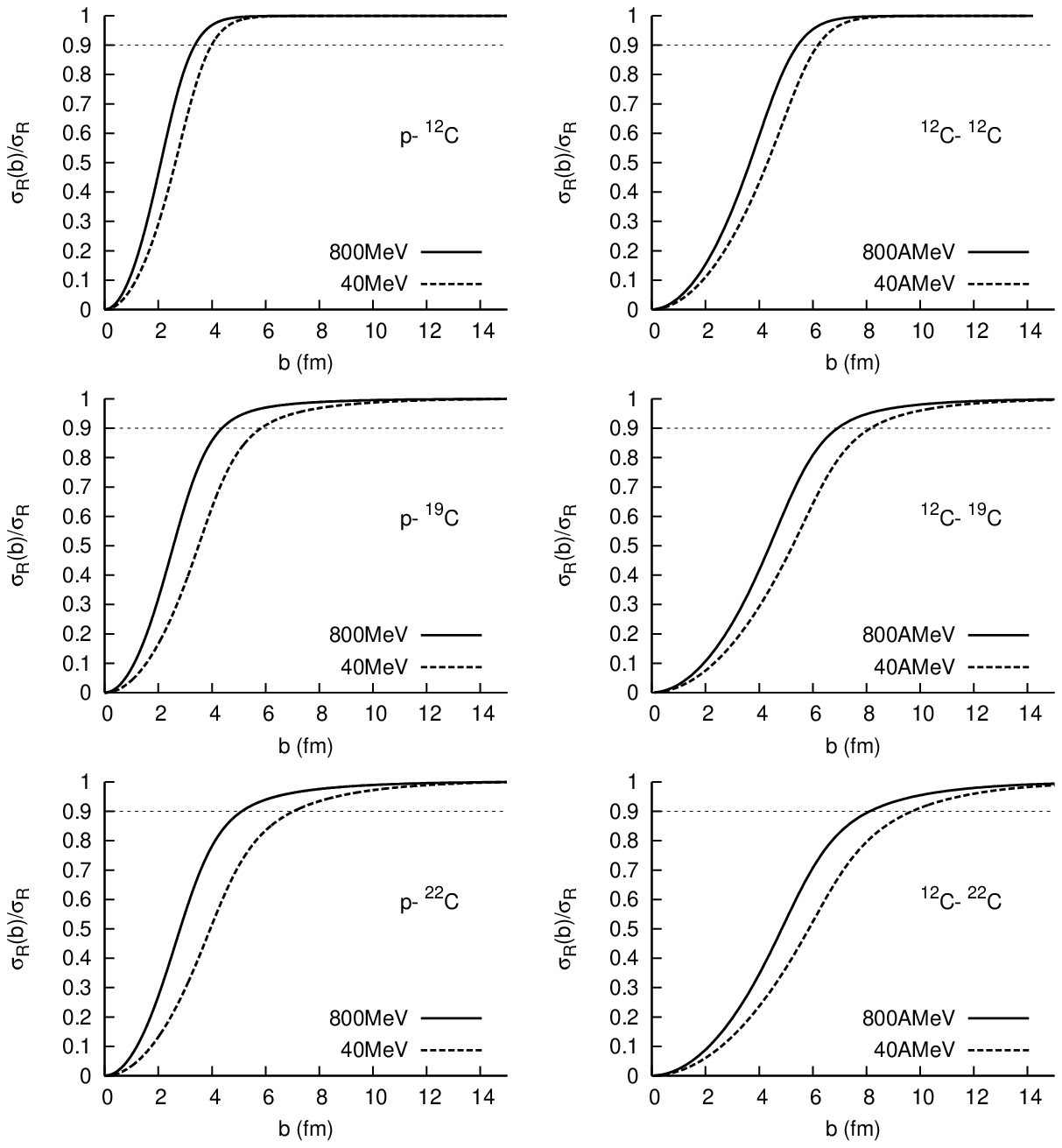,width=13cm,height=13cm}
\caption{Comparison of the calculated reaction cross section 
at a given 
impact parameter, $b$,
in ratio to the total reaction cross section 
between proton-$^{12,19,22}$C and $^{12}$C-$^{12,19,22}$C. 
The solid curve is the calculation at 800 MeV, while the 
dashed curve is at 40 MeV. 
The dotted line indicates $90\,\%$ of $\sigma_{\rm R}$.
}
\label{reac-b}
\end{figure*}

In order to compare the sensitivity of the proton and 
carbon probes to the nuclear surface, 
we plot, in Fig.~\ref{reac-b}, $\sigma_{\rm R}(b)/\sigma_{\rm R}$ 
for three nuclei, $^{12}$C, $^{19}$C and $^{22}$C, 
of different features 
as a function of the impact parameter, $b$, 
where $\sigma_{\rm R}(b)$ is defined similarly 
to Eq.~(\ref{reaccs}) but the upper limit of the integration 
is limited to $b$. The nucleus $^{12}$C is a stable 
nucleus which has almost the same proton and neutron distributions,  
$^{19}$C is a good example of one-neutron halo nucleus, whereas 
$^{22}$C is a two-neutron halo nucleus with a long neutron tail. 

As one can see from the figures, for each case, 
the major contribution comes from the surface region, 
which supports the above discussion. 
As a rough estimate of the extent to which 
the surface region is probed, 
we may use an impact parameter at which $\sigma_{\rm R}(b)$ 
reaches 90$\,\%$ of $\sigma_{\rm R}$. Then we take the difference 
of such impact parameters, $\Delta b$, between 40 and 800 MeV 
incident energies. 
The increase of $\Delta b$ for the change 
of incident energy from 800 to 40 MeV 
is understood from the fact that the $pn$ interaction becomes 
longer-ranged and stronger, which is reflected in the 
energy-dependence of $\beta_{pn}$ and $\sigma_{pn}^{\rm tot}$. 

First we focus on the reaction cross sections for the proton 
target. The $\Delta b$ value increases from 0.6, 1.5 to 1.9 fm as 
the neutron density becomes more widely distributed for 
$^{12}$C, $^{19}$C and $^{22}$C, respectively. This suggests that 
the proton target can probe the density distribution 
near the surface up to further distances 
as the interaction range increases. The 
corresponding $\Delta b$ value for the $^{12}$C target case is 
0.7, 1.3 and 1.6 fm for $^{12}$C, $^{19}$C and $^{22}$C, 
respectively. 
Comparing $\Delta b$ values for the proton targets with 
those for $^{12}$C targets, 
we can conclude that the $^{12}$C target can probe the surface 
region equally to the proton target but is disadvantageous to  
probe the remote surface region of 
the spatially extended neutron distribution, such as $^{22}$C, 
compared to the proton. 
This is due to the fact that the proton and neutron 
distributions in $^{12}$C are very similar and that the $nn (pp)$ 
interaction is shorter-ranged and weaker than the $pn$ interaction.

It would be possible to probe 
the outer region of the density distribution 
by the proton target especially at lower energy, 
but, at very low energy, we have to note that the long 
wavelength of the proton 
leads to a low resolution to the resultant density 
distributions, 
which may prevent us from studying minute structures of 
the outer density in detail.

\section{summary}

We have made a systematic analysis of 
the total reaction cross sections 
for the carbon isotopes incident on 
a proton target for wide energy range, 
in comparison with the features of a carbon target.
We have predicted the reaction cross sections 
especially at 40 MeV where the experimental data 
have been measured at RIKEN.

We have formulated this problem using the Glauber theory.
The inputs are the parameters of 
nucleon-nucleon profile functions and 
the wave functions (densities) of carbon isotopes. 
The parameters of the nucleon-nucleon scattering 
are determined from the available experimental data. 
The densities are generated using the wave functions of 
the Slater determinant, 
which we used in our previous work \cite{hsbk}.
To go beyond that, we use 
a core+$n$ model for odd nuclei 
and a core+$n$+$n$ model for $^{16,22}$C nuclei. 

Having treated the interactions of proton-proton 
and proton-neutron separately, 
we have shown that the optical limit approximation 
of the Glauber theory 
gives almost the same results as the 
few-body calculation for the proton-nucleus 
reaction cross section over all the energy range used here.

For $^{22}$C, we generate several densities 
which are constructed from the wave functions 
giving different 
separation energies of 0.489, 0.361, 0.232 
and 0.122 MeV for the last two neutrons. 
All of them lie within the error bar of the 
experimental value, 0.423$\pm$1.140 MeV. 
The reaction cross sections calculated using these densities  
are 957, 969, 985 and 1005 mb, respectively, 
for proton-$^{22}$C at 40 MeV. 
Since the preliminary data of proton-$^{22}$C 
reaction cross section has been reported 
to be around 1000 mb with a large uncertainty, 
all of our predictions are consistent with the data, 
but the larger two values, 985 and 1005 mb, would be favorable. 
If so, the data may suggest very small $S_{2n}$.

At around 100 MeV, the values of $r_{\rm BS}$ defined by 
Eq.~(\ref{rbs}) happen to agree 
reasonably well with $r_m$ listed in Table~\ref{radii}, 
which suggests that we may empirically access to the rms 
nuclear matter radii of carbon isotopes 
just by measuring $\sigma_{\rm R}$ at 100 MeV. 

We have found a parameter free new relation, Eq.~(\ref{emp}).  
It helps us to predict reaction 
cross sections for various isotopes at a given energy 
if the reaction cross section value of some stable isotope is 
available.  

Finally, we have made simple estimates for the contribution 
of the neutron and the proton to the total reaction cross 
sections.  
The major contribution to $\sigma_{\rm R}$ comes from 
the surface region. 
Moreover, we have pointed out that a proton 
target can probe the surface region of the neutron-rich nuclei 
better than a $^{12}$C target especially at lower incident 
energy. 

\vspace{5mm}

We acknowledge T. Motobayashi for his encouragement
during the course of this work. 
W. H. is a Research Fellow 
of the Japan Society for the Promotion of Science for 
Young Scientists.
A. K. would like to thank K. Iida, K. Oyamatsu, and M. Takashina for 
useful comments and helpful discussions. 
This work was in part supported by a Grant for Promotion of Niigata 
University Research Projects (2005--2007), 
and a Grant-in Aid for 
Scientific Research for Young Scientists (No. 19$\cdot$3978). 
One of the authors (Y.S.) thanks the JSPS core-to-core program, 
Exotic Femto Systems.

\section*{Appendix: $NN$ Scattering Amplitudes}

Here we discuss the parameterizations of 
nucleon-nucleon scattering amplitudes.
In the text, 
we parameterize it with a single Gaussian. 
We show here that the parameterization with 
double Gaussians gives numerically
almost the same results for the reaction cross sections 
as the single Gaussian, and validates our use 
of the single Gaussian prescription. 

We only show the case of $pn$ scattering, 
because its contribution is more important 
for the neutron-rich isotopes than $pp$ scattering, 
and also because, as we have discussed in Sec.~V, 
the $pn$ reaction dominates the proton-nucleus reaction 
cross sections especially at energies less than 100 $A$MeV.

\begin{figure*}[t]
\epsfig{file=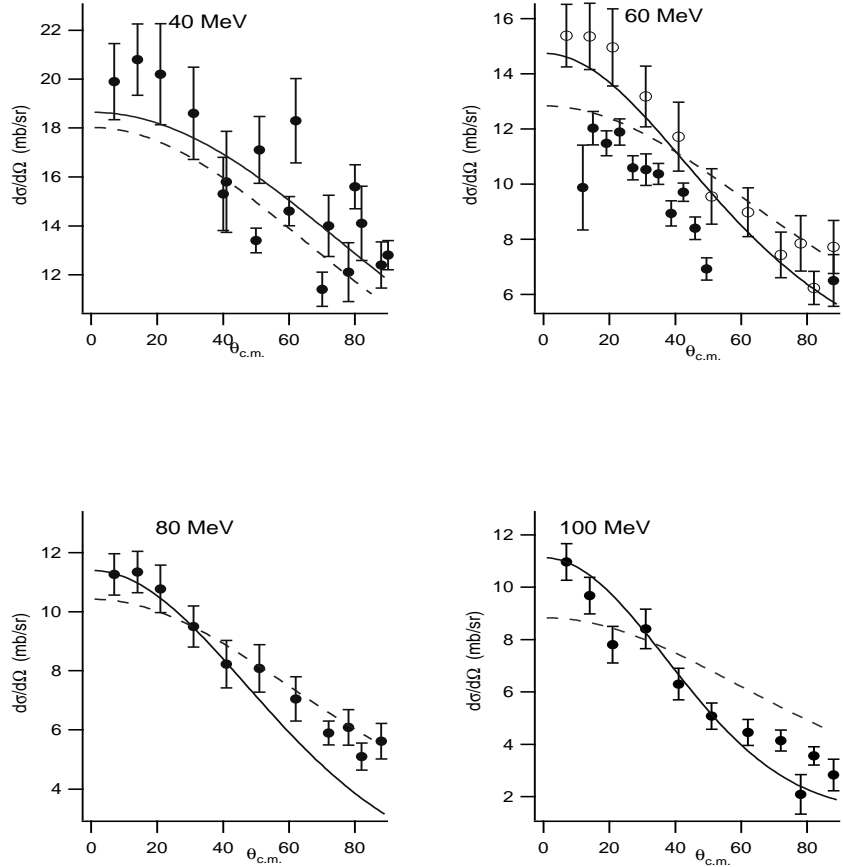,width=12cm,height=12cm}
\caption{The differential cross sections of
the $pn$ elastic scattering. 
The solid curves denote the fitting using double Gaussians, 
as explained in the text, while 
the dashed curves the results with the 
parameters listed in Table~\ref{parameter}. 
The experimental data are taken from Ref.~\cite{scanlon}.}
\label{pn}
\end{figure*}

In Fig.~\ref{pn}, the numerical results of
the $pn$ elastic scattering differential cross sections
calculated using the parameters of Ref.~\cite{hostachy} 
are compared with the data. 
The numerical results of the single Gaussian 
are shown by the dashed curve. 
The agreement of the results with the data is reasonable, 
but not perfect especially in the forward direction.

Although the agreement with the data 
of the $pn$ elastic differential cross section 
is imperfect, the expression of the single Gaussian 
well reproduces the reaction cross sections as in Table~\ref{p12c}.

For comparison, 
we perform fittings for the $pn$ 
elastic scattering data in 
the energy range from 40 MeV to 100 MeV, 
using the parameterization of double Gaussians.
For this case, the profile function, $\Gamma_{pN}^{\rm D}$, 
for $pp$ and $pn$ scatterings, is parameterized in the form
\begin{equation}
  \Gamma_{pN}^{\rm D}({\bm b}) =
  \frac{1-i\alpha_{1}}{4\pi\beta_{1}}\,\,
  \sigma_{1}\, {\rm e}^{-{\bm b}^2 /(2\beta_{1}) }+
\frac{1-i\alpha_{2}}{4\pi\beta_{2}}\,\,
  \sigma_{2}\, {\rm e}^{-{\bm b}^2 /(2\beta_{2}) },
\label{gfn2}
\end{equation}
where, $\sigma_{1}$, $\alpha_{1}$, $\beta_{1}$, 
$\sigma_{2}$, $\alpha_{2}$ and $\beta_{2}$ are fitting parameters 
determined from the requirement: 
1)~The optical theorem is satisfied. 
2)~The ratio of the real to the imaginary part of 
the $pn$($pp$) scattering amplitude in the forward direction 
reproduces the experimental values. 
3)~The total elastic scattering cross section is equal 
to the total cross section. 
4)~The elastic scattering differential cross sections 
are reproduced. 

The fitting results using the double Gaussians
are displayed by the solid curves in Fig.~\ref{pn}.
The two sets of experimental data shown as 60 MeV 
in Fig.~\ref{pn} are at 62 MeV (open circle)
and 63 MeV (thick dot).
As for the differential cross sections, 
it seems that the double Gaussians give better results.

In Table~\ref{p12c}, we compare 
the numerical results of proton-$^{12}$C total 
reaction cross sections. 
The calculations using the double Gaussians are the fit, 
while those using the single Gaussian are obtained 
by the use of the parameters of Ref.~\cite{hostachy}.
The parameters of $pp$ scattering are kept fixed. 
The experimental data are also shown in the table.
The difference between the results of 
the reaction cross sections using these two parameterizations 
is a few $\%$ except at 60 MeV, 
but around this energy the data scatter widely and 
the difference between the numerical results and the data 
is not serious. 

\begin{table}[h]
\caption{Total reaction cross sections of proton-$^{12}$C in mb 
calculated using the parameters given in Table~\ref{parameter}, 
and that determined from two Gaussian fitting for 
$pn$ elastic scattering data. 
The experimental data are taken from Refs.~\cite{carlson,auce}.}
\label{p12c}
\begin{center}
\begin{tabular}{cccc}
\hline\hline
$E$ (MeV)    & Table~\ref{parameter}& present fit & Exp. \\
\hline
40  & 432 &  416 & 371(11), 405(38) \\
60  & 359 &  387 & 310 (13)   \\       
80  & 314 &  320 & 279 (10)    \\       
100 & 284 &  294 & 275 (21)     \\       
\hline\hline
\end{tabular}
\end{center}
\end{table}

Thus, we conclude that the parameters listed 
in Table~\ref{parameter} works fairly well for the 
reaction cross sections. For simplicity, 
we adopt the parameterization with the single Gaussian
throughout this paper.

\end{document}